\documentclass[10pt,journal,cspaper,compsoc]{IEEEtran}
\ifCLASSOPTIONcompsoc
\else
\fi
\usepackage{graphicx}

\usepackage{subfigure}
\usepackage{algorithm}
\usepackage[noend]{algorithmic}
\usepackage{xspace}
\usepackage{latexsym}

\ifCLASSINFOpdf
\else
\fi
\begin{document}

\newtheorem{theorem}{Theorem}
\newtheorem{example}{Example}
\newtheorem{property}{Property}
\newtheorem{definition}{Definition}

\title{Quasi-SLCA based Keyword Query Processing over Probabilistic XML Data}

\author{Jianxin Li, Chengfei Liu, Rui Zhou and Jeffrey Xu Yu,~\IEEEmembership{Member,~IEEE,}
\IEEEcompsocitemizethanks{
\IEEEcompsocthanksitem Jianxin Li, Chengfei Liu and Rui Zhou are with the Faculty of Information \& Technology, Swinburne University of Technology, Australia. \{jianxinli, cliu, rzhou\}@swin.edu.au \protect\\

\IEEEcompsocthanksitem Jeffrey Xu Yu is with the Department of Systems Engineering \& Engineering Management, The Chinese University of Hong Kong, China. yu@se.cuhk.edu.hk \protect\\

}
\thanks{}}

\IEEEcompsoctitleabstractindextext{%
\begin{abstract}

The probabilistic threshold query is one of the most common queries in uncertain databases, where a result satisfying the query must be also with probability meeting the threshold requirement.
In this paper, we investigate probabilistic threshold keyword queries (PrTKQ) over XML data, which is not studied before. 
We first introduce the notion of quasi-SLCA and use it to represent results for a PrTKQ with the consideration of possible world semantics. 
Then we design a probabilistic inverted (PI) index that can be used to quickly return the qualified answers and filter out the unqualified ones based on our proposed lower/upper bounds. 
After that, we propose two efficient and comparable algorithms: Baseline Algorithm and PI index-based Algorithm. 
To accelerate the performance of algorithms, we also utilize probability density function. 
An empirical study using real and synthetic data sets has verified the effectiveness and the efficiency of our approaches.

\end{abstract}

\begin{keywords}
Probabilistic XML, Threshold Keyword Query, Probabilistic Index.
\end{keywords}}

\maketitle

\IEEEdisplaynotcompsoctitleabstractindextext

%
\IEEEpeerreviewmaketitle

\section{Introduction}
Uncertainty is widespread in many web applications, such as information extraction, information integration, web data mining, etc. 
In uncertain database, probabilistic threshold queries have been studied extensively where all results satisfying the queries with probabilities equal to or larger than the given threshold values are returned \cite{DBLP:journals/sigmod/ChengP03,DBLP:conf/vldb/ChengXPSV04,DBLP:conf/vldb/TaoCXNKP05,DBLP:conf/sigmod/HuaPZL08,DBLP:conf/sigmod/QiJSP10}. However, all of these works were studied based on uncertain relational data model. 
Because the flexibility of XML data model allows a natural representation of uncertain data, uncertain XML data management has become an important issue and lots of works have been done recently. For example, many probabilistic XML data models were designed and analyzed  \cite{DBLP:conf/pods/SenellartA07,DBLP:conf/vldb/NiermanJ02,DBLP:conf/icde/HungGS03,DBLP:conf/icde/KeulenKA05,DBLP:conf/edbt/AbiteboulS06}. 
Based on different data models, query evaluation  \cite{DBLP:conf/vldb/NiermanJ02,DBLP:conf/edbt/AbiteboulS06,DBLP:conf/vldb/KimelfeldS07,DBLP:conf/sigmod/KimelfeldKS08,DBLP:conf/edbt/ChangYQ09}, 
algebraic manipulation \cite{DBLP:conf/icde/HungGS03} and updates \cite{DBLP:conf/pods/SenellartA07,DBLP:conf/edbt/AbiteboulS06} were studied.  
However, most of these works concentrated on structured query processing, e.g., twig queries. 
In this paper, we propose and address a new interesting and challenging problem of \textit{Pr}obabilistic \textit{T}hreshold \textit{K}eyword \textit{Q}uery (PrTKQ) over uncertain XML databases based on quasi-SLCA semantics, which is not studied before as far as we know.




In general, an XML document could be viewed
as a rooted tree, where each node represents an element
or contents. XIRQL~\cite{DBLP:conf/sigir/FuhrG01} supports keyword search in XML based on structured queries. However, users may not have the knowledge of the structure of XML data or the query language. As such, supporting pure keyword search in XML has attracted extensive research. The LCA-based approaches will
identify the LCA node first, which contains every
keyword under its subtree at least once \cite{DBLP:conf/vldb/CohenMKS03,DBLP:conf/sigmod/GuoSBS03,DBLP:conf/www/KoloniariP10,DBLP:conf/cikm/LiFWZ07,DBLP:conf/vldb/LiYJ04,DBLP:conf/sigmod/XuP05,DBLP:conf/edbt/ZhouLL10}.  
Since the LCA nodes sometimes are not
very specific to users' query, Xu and Papakonstantinou
\cite{DBLP:conf/sigmod/XuP05} proposed the concept of SLCA (smallest
lowest common ancestor), where a node $v$ is regarded as an SLCA if (a) the subtree rooted at the node $v$, denoted as $T_{sub}(v)$, contains all the keywords, and (b) there does not exist a descendant node $v'$ of $v$ such that $T_{sub}(v')$ contains all the keywords. In other words, if a node is an SLCA, then its ancestors will be definitely excluded from being SLCAs. The SLCA semantics of model keyword search result on a deterministic XML tree are also applied \cite{DBLP:conf/www/SunCG07,DBLP:conf/sigmod/GuoSBS03,DBLP:conf/vldb/LiYJ04}. 


Based on the SLCA semantics, \cite{DBLP:conf/icde/LiLZW11} discussed top-$k$ keyword search over a probabilistic XML document. 
Given a keyword query $q$ and a probabilistic XML document (PrXML), \cite{DBLP:conf/icde/LiLZW11} returned the top $k$ most relevant SLCA results (PrSLCAs) based on their probabilities.
Different from the SLCA semantics over deterministic XML documents, a node $v$ being a PrSLCA can only exclude its ancestors from being PrSLCAs by a probability.
This probability can be calculated by aggregating the probabilities of the deterministic documents (called possible worlds) $W$ implied in the PrXML where $v$ is an SLCA in each deterministic document $\in W$. 
 

However, it is not suitable to 
directly utilize the PrSLCA semantics for evaluating PrTKQs because the PrSLCA semantics are too strong. 
In some applications, users tend to be confident with the results to be searched, so relatively high probability threshold values may be given. Consequently, it is very likely that no qualified PrSLCA results will be returned.
 To solve this problem, we propose and utilize a so-called \textit{quasi-SLCA} semantics to define the results of a PrTKQ by \textit{relaxing} the semantics of PrSLCA with regards to a given threshold value, 
i.e., besides the probability of $v$ being a PrSLCA in PrXML, the probability of a node $v$ being a quasi-SLCA in PrXML may also count the probability of $v'$s descendants being PrSLCAs in PrXML if their probabilities are below the specified threshold value. In other words, a node $v$ being a quasi-SLCA will exclude its ancestors from being quasi-SLCAs by a probability only when this probability is no less than the given threshold; otherwise, this probability will be included for  contributing to its ancestors. This is different from the PrSLCA semantics that excludes the probability contribution from child nodes.

\begin{figure}[htbp]
\centering
\includegraphics[scale=0.55]{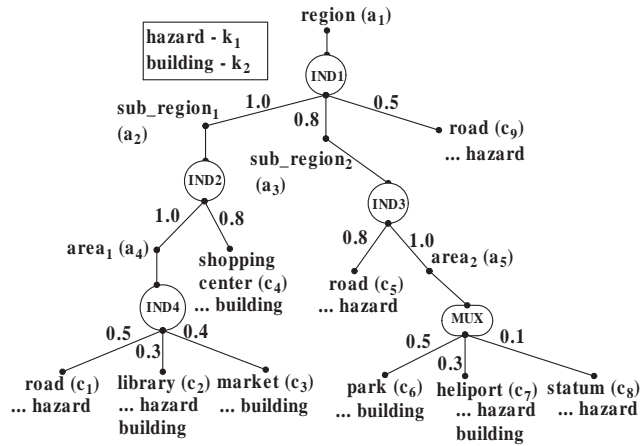}
\caption{A probabilistic XML data tree}
\label{fig:pxml}
\end{figure}


\begin{example}\label{exam:motivation1}
Consider an aircraft-monitored battlefield application, where the useful information will be taken as Aerial photographies. Through analysing the photographies, we can extract the possible objects (e.g., road, factory, airport, etc.) and attach some text description to them with probabilities, which can be stored in the format of PrXML. Figure~\ref{fig:pxml} is a snapshot of an aircraft-monitored battlefield XML data.
By issuing a keyword query $\{hazard, building\}$, a military department  would find the potential areas containing hazard buildings above a probability threshold. 

Based on the semantics of PrSLCA, any of the nodes $library$ (probability = 0.3), $area_1$( = 0.14), $sub\_region_1$( = 0.168), $heliport$( = 0.24), $sub\_region_2$( = 0.32) and $region$( = 0.088) can become an PrSLCA result. The detailed procedure of calculating the probabilities of results will be shown later. As we know, the users generally specify a threshold value $\sigma$ as the confidence score with their issued query, e.g., $\sigma = 0.40$ representing that the users prefer to see the answers with their probabilities up to 0.40. In this condition, no results can be returned to the users.     

However, from Figure~\ref{fig:pxml}, we can see that if the probabilities of $library$ and $area_2$ could contribute to their parent nodes, $area_1$ and $sub\_region_2$ would become quasi-SLCA results. Unfortunately, the PrSLCA semantics exclude them from being results. This motivates us to relax the PrSLCA semantics to the quasi-SLCA semantics. According to the quasi-SLCA semantics, the probabilities of $area_1$ and $sub\_region_2$ being the quasi-SLCA results are 0.44 and 0.56 with the contributions of their child nodes $library$ and $area_2$, respectively. As such, $area_1$ and $sub\_region_2$ are deemed as the interesting places to be returned.
  \end{example}  

 Given a PrTKQ, our problem is to quickly compute all the quasi-SLCA nodes
with their probabilities meeting the threshold requirement.  
For users issuing PrTKQs, they generally expect to see the complete quasi-SLCA answer set as early as possible and do not need to know the accurate probability of each answer, which motivates us to design a \textit{P}robabilistic \textit{I}nverted (PI) index and PI-based efficient algorithm for quickly identifying quasi-SLCA result candidates.  

We summarize the contributions of this paper as follows:
\begin{itemize}

\item Based on our proposed quasi-SLCA result definition, we study probabilistic threshold keyword query over uncertain XML data, which satisfies the possible world semantics. To the best of our knowledge, this problem has not been studied before.

\item We design a probabilistic inverted (PI) index that can quickly compute the lower bound and upper bound for a threshold keyword query, by which lots of unqualified nodes can be pruned and qualified nodes can be returned as early as possible. To keep the effectiveness of pruning, the probability density function is employed based on the assumption of Gaussian distribution.

\item We propose two algorithms, a comparable baseline algorithm and a PI-based Algorithm, to efficiently find all the quasi-SLCA results meeting the threshold requirement.  

\item Experimental evaluation has demonstrated the efficiency and effectiveness of the proposed approaches.

\end{itemize}


The rest of this paper is organized as follows. In Section~\ref{sec:problemdefinition}, we introduce the probabilistic XML model and the problem definition of probabilistic threshold keyword query. Section~\ref{sec:overview} shows the procedure of efficiently finding quasi-SLCA results using an example. Section~\ref{sec:index} first presents the data structure of PI index, discusses the basic building operations and pruning techniques of PI index, and provides the building algorithm of PI index. In Section~\ref{sec:algorithms}, we propose a comparable baseline algorithm and a PI-based algorithm to find the qualified quasi-SLCA results. We report the experimental results in Section~\ref{sec:experimentalresults}. Section~\ref{sec:relatedwork} discusses related works and Section~\ref{sec:conclusions} concludes the paper.

\section{Probabilistic Data Model and \\Problem Definition}\label{sec:problemdefinition}

\textbf{Probabilistic Data Model:}
A PrXML document defines a probability distribution over a space of deterministic XML documents. 
Each deterministic document belonging to this space is called a possible world.
A PrXML document represented as a labelled tree has \textit{ordinary} and \textit{distributional} nodes.
Ordinary nodes are regular XML nodes and they may appear in deterministic documents, 
while distributional nodes are only used for defining the probabilistic process of generating deterministic documents 
and they do not occur in those documents. 

In this paper, we adopt a popular probabilistic XML model, PrXML$^{\{ind, mux\}}$ \cite{DBLP:conf/sigmod/KimelfeldKS08,DBLP:conf/icde/LiLZW11}, which was first discussed in \cite{DBLP:conf/vldb/NiermanJ02}. In this model, a PrXML document is considered as a labelled tree where distributional nodes have two types, IND and MUX. An IND node has children that are \emph{independent} of each other, while the children of a MUX node are \emph{mutually-exclusive}, that is, at most one child can exist in a random instance document (called a \emph{possible world}). A real number from (0,1] is attached on each edge in the XML tree, indicating the conditional probability that the child node will appear under the parent node given the existence of the parent node. An example of a PrXML document is given in Fig.~\ref{fig:pxml}. Unweighted edges have 1 as the default conditional probability.



\textbf{The Semantics of PrSLCA in PrXML:}
According to the semantics of possible worlds, the global probability of a node $v$ being a PrSLCA with regard to a given query $q$ in the possible worlds is defined as follows:
\begin{equation}
Pr^{G}_{slca}(q, v) = \sum_{i=1}^{m} \{Pr(w_i)|slca(q,v,w_i)=true\}
\label{eq:defineprob}
\end{equation}   
where $w_1, \ldots, w_m$ denotes the possible worlds implied by 
 $slca(q,v,w_i)=true$ indicates that $v$ is an SLCA in the possible world $w_i$ for the query $q$. $Pr(w_i)$ is the existence probability of the possible world $w_i$. The symbol $G$ means $Pr^{G}_{slca}(q, v)$ is the \textit{global} probability of a node $v$ being an SLCA w.r.t. $q$ in all possible worlds.

\begin{figure}[ht]
\centering
\includegraphics[scale=0.45]{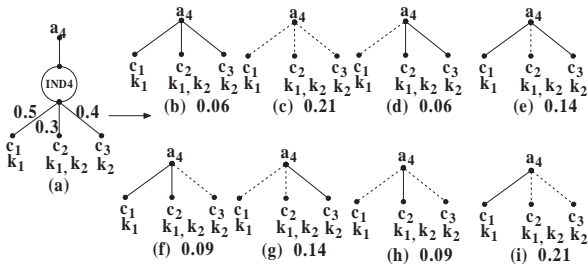}
\caption{A small PrXML and its possible worlds}
\label{fig:smallpxml}
\end{figure}

\begin{example}\label{exam:smallpxml}
Consider a small PrXML in Figure~\ref{fig:smallpxml}.a and all generated possible worlds in Figure~\ref{fig:smallpxml}.\{b,c,d,e,f,g,h,i\} where the solid line represents the existence of the edge while the dashed line represents the absence of the edge. Given a possible world, we can compute its global probability based on the existence/absence of the edges in the possible world, e.g., $Pr(w_d) = (1 - 0.5) * 0.3 * 0.4 = 0.06$. 

Given a keyword query $q=\{k_1, k_2\}$, we can compute the global probability of $c_2$ being a PrSLCA w.r.t. $q$ by using $Pr^{G}_{slca}(q, c_2) = Pr(w_b) + Pr(w_d) + Pr(w_f) + Pr(w_h) = 0.06 + 0.06 + 0.09 + 0.09 = 0.30$. 
Similarly, we have the global probability of $a_4$ being a PrSLCA w.r.t. $q$ by using $Pr^{G}_{slca}(q, a_4) = Pr(w_e) = 0.14$.
\end{example}

 \textbf{The Semantics of quasi-SLCA in PrXML:}

\begin{definition}\label{def:quasislca}
\textbf{Quasi-SLCA:} Given a keyword query $q$ and a threshold value $\sigma$, 
a node $v$ is called a quasi-SLCA if and only if  
(1) $v$ or its descendants are SLCAs in a set $W$ of possible worlds;
(2) the aggregated probability of $v$ and its descendants to be SLCAs in $W$ is no less than $\sigma$;
(3) no descendant nodes of $v$ satisfy both of the above conditions in any set of possible worlds that overlaps with $W$. 
\end{definition}

In other words, if a descendant node $v_d$ of $v$ is a quasi-SLCA, then the probability of $v_d$ has to be excluded from the probability of $v$ being a quasi-SLCA. It means that the set of possible worlds that $v_d$ appears does not overlap with the set of possible worlds that $v$ or its other descendants appear.

Given a query $q$, we can compute $Pr^{L}_{slca}(q, v)$ in a bottom-up manner, where $Pr^{L}_{slca}(q, v)$ stands for the local probability for $v$ being an SLCA in the probabilistic subtree rooted at $v$. 
For example, $a_4$ in Figure~\ref{fig:smallpxml}(a) is a subtree of Figure~\ref{fig:pxml}. 
$Pr^{L}_{slca}(q, a_4)$ can be used to compute the PrSLCA probability of $a_2$ and $a_1$.
From $Pr^{L}_{slca}(q, v)$, we can easily get $Pr^{G}_{slca}(q,v)$ by 
$Pr^{G}_{slca}(q, v)$ = $Pr(path_{r \rightarrow v}) \times Pr^{L}_{slca}(q, v)$
where $Pr(path_{r \rightarrow v})$ indicates the existence probability of $v$ in the possible worlds. It can be computed by multiplying the conditional probabilities along the path from the root $r$ to $v$. 

Now, we define quasi-SLCA based on PrSLCA and the parent-child relationship.
For an IND node $v$, we have: 
\begin{equation}\label{equ:quasiind}
\begin{array}{l}
Pr^{G}_{quasi-slca}(q,v) = Pr^{G}_{slca}(q,v) + Pr(path_{r \rightarrow v}) \times \\
              (1 - \prod_{v' \in child(v) \wedge v'\notin V_{quasi}}(1-Pr^{L}_{slca}(q,v')))
\end{array}
\end{equation}
where the child node $v'$ of $v$ is an SLCA node, but not a quasi-SLCA node. 

For MUX node $v$, we have:
\begin{equation}\label{equ:quasimux}
\begin{array}{l}
Pr^{G}_{quasi-slca}(q,v) = Pr^{G}_{slca}(q,v) + Pr(path_{r \rightarrow v}) \times \\ 
\sum \{Pr^{L}_{slca}(q,v')|v' \in child(v) \wedge v'\notin V_{quasi}\}
\end{array}
\end{equation}

Note, IND or MUX nodes are normally not allowed to be SLCA result nodes because they are only distributional nodes. 
As such, for the above IND or MUX node $v$, we may use its parent node $v_p$ (with $v$ as a sole child) to represent the SLCA result node. 





\begin{example}\label{exam:quasislca}
Let's consider Example~\ref{exam:smallpxml} again.
First assume the specified threshold value is 0.40, then the global probability of $a_4$ being a quasi-SLCA result can be calculated by using 
$Pr^{G}_{quasi-slca}(q, a_4)$ = $Pr^{G}_{slca}(q, a_4)$ + $Pr(path_{r \rightarrow a_4})$ * $(1 - (1 - Pr^{L}_{slca}(q, c_2)))$ = 0.14 + 0.30 = 0.44 because child $c_2$ is an SLCA node but not a quasi-SLCA node w.r.t. the given threshold. So $c_2$'s SLCA probability contributes to its parent node $a_4$.
If the threshold is decreased to 0.30, then $c_2$ will be taken as a qualified quasi-SLCA result and will not contribute to $a_4$. 
In this case, $a_4$ cannot become a quasi-SLCA result because $Pr^{G}_{quasi-slca}(q, a_4)$ = $Pr^{G}_{slca}(q, a_4)$ = 0.14 $<$ 0.30. If the threshold is further decreased to 0.14, both $c_2$ and $a_4$ are qualified quasi-SLCA results. 
\end{example}

\begin{definition}
\textbf{\underline{Pr}obabilistic \underline{T}hreshold \underline{K}eyword \underline{Q}uery:} (PrTKQ) Given a keyword query $q$ and a threshold $\sigma$, the results of $q$ over a probabilistic XML data $T$ is a set $R$ of quasi-SLCA nodes with their probabilities equal to or larger than 
$\sigma$, i.e., $Pr^{G}_{quasi-slca}(q, v) \geq \sigma$ for $\forall v \in R$.
\end{definition}



In this work, we are interested in how to efficiently compute the quasi-SLCA answer set for a PrTKQ over a probabilistic XML data.

\section{Overview of this Work}\label{sec:overview}

A naive method to answer a PrTKQ is to enumerate all possible worlds and apply
the query to each possible world. Then, we can compute the overall probability
of each quasi-SLCA result and return the results meeting the probability threshold. However, the naive method is inefficient due to the huge number of possible worlds over a probabilistic XML data. Another method is to extend the work in \cite{DBLP:conf/icde/LiLZW11} to compute the probabilities of quasi-SLCA candidates. Although it is much more efficient than the naive method, it needs to scan the keyword node lists and calculate the keyword distributions for all relevant nodes. Therefore, that motivates our development of efficient algorithms which not only avoids generating possible worlds, but also prunes more unqualified nodes.

\begin{figure}[ht]
\centering
\includegraphics[scale=0.5]{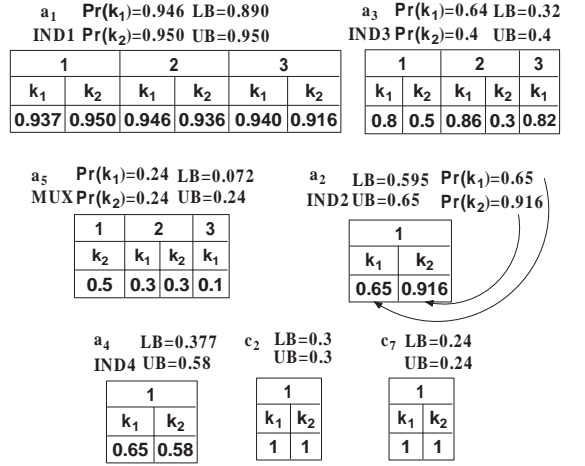}
\caption{PI index and Lower/Upper Bound for a query $\{k_1, k_2\}$ over the given PrXML}
\label{fig:bound}
\end{figure}

To accelerate query evaluation, in this paper we propose a prune-based probabilistic threshold keyword query algorithm, which determines the qualified results and filters the unqualified candidates by using off-line computed probability information. To do this, we need to first calculate the probability of each possible query term within a node, which is stored as an off-line computed probabilistic index. 
Within a node, any two of its contained terms may appear in the IND or MUX ways. To precisely differentiate IND and MUX, we utilize different parts to represent the probabilities of possible query terms appearing in MUX way, while the terms in each part hold IND relationships. In other words, the different parts of terms in a node are mutual-exclusive (MUX), e.g., $a_1$ and $a_5$ in Figure~\ref{fig:bound} consists of three parts.

Given a keyword query and a threshold value, we first load the corresponding off-line computed probabilistic index w.r.t. the keyword query and then on-the-fly calculate the range of probabilities of a node being a result of the keyword query using the pre-computed probabilistic index in a bottom-up strategy. 
Here, the range of probabilities can be represented by two boundary values: lower bound and upper bound. By comparing the lower/upper bounds of candidates, the qualified results can be efficiently identified.

The followed two examples briefly demonstrate how we calculate the lower/upper bounds based on a given keyword query and the off-line computed probabilistic index, and how we apply the on-line computed lower/upper bounds to prune the unqualified candidates and determine the qualified ones. 

 Figure~\ref{fig:bound} shows the lower/upper bounds of each node in Figure~\ref{fig:pxml} where the probability of each individual term is calculated offline while the lower/upper bounds are computed on-the-fly based on the given query keywords.
Let's first introduce the related concepts briefly: 
the probability of a term in a node represents the total local probability of the term appearing in all possible worlds to be generated for the probabilistic subtree rooted at the node, e.g., $Pr(k_1, a_2)$ = 0.65 and $Pr(k_2, a_2)$ = 0.916; 
 the lower bound value represents the minimal total local probability of the given query keywords appearing in all the possible worlds w.r.t. the probabilistic subtree, e.g., LB($k_1k_2$, $a_2$)=0.65*0.916=0.595;
 the upper bound value represents the maximal total local probability of the given query keywords appearing in all the possible worlds  w.r.t. the probabilistic subtree because the keywords may be independent or co-occur, e.g., UB($k_1k_2$, $a_2$) = min\{0.65, 0.916\} = 0.65 no matter whether they are independent. 
 By multiplying the path probability, the local probability can be transformed into the global probability. 
 For the nodes containing MUX semantics, we group the probabilities of its terms into different parts, any two of which are mutually-exclusive as shown in $a_1$, $a_3$ and $a_5$ in Figure~\ref{fig:bound}.
  The details of computing the lower/upper bounds for the IND and MUX semantics in the following section. 

 

\begin{example}\label{exam:motivation}
 Consider a PrTKQ $\{k_1, k_2\}$ with $\sigma$=0.40 again.  
 $a_5$, $c_2$ and $c_7$ can be pruned directly without calculation because their upper bounds are all lower than 0.40.
 We need to check the rest nodes $a_1$, $a_2$, $a_3$ and $a_4$.
 For $a_4$, after computation, the probability of $a_4$ being a quasi-SLCA result is 0.44, which is larger than the specified threshold value 0.40, so $a_4$ will be taken as a result. After that, the result of $a_4$ can be used to update the lower bound and upper bound of $a_2$, (LB=0.595, UB=0.65) $\rightarrow$ (LB=0.155, UB=0.21). As a consequence, $a_2$ should be filtered due to $UB(a_2)=0.21 < \sigma = 0.40$.
Similarly, $a_3$ can be computed and selected as a result because its probability is 0.56. 
Since $a_3$ and $a_4$ having been the quasi-SLCA results, the bounds of $a_1$ can be updated as (LB=0.890, UB=0.950) $\rightarrow$ (LB=0.136, UB=0.196). As such, $a_1$ can be pruned because its upper bound is lower than 0.40. From this  example, we can find that many answers can be pruned or returned without the need to know their accurate probabilities, and the effectiveness of pruning would be accelerated greatly with the increase of users' search confidence.
\end{example}

As an acute reader, you may find that we have to compute the probability of $a_4$ being a 
quasi-SLCA because it cannot determine whether or not $a_4$ is a qualified result to be output only based on its lower/upper bound values. To exactly calculate the probability of $a_4$ being a 
quasi-SLCA, we have to access its child/descendant nodes, e.g., $c_1, c_2, c_3$, although $c_2$ has been recognized as a pruned node before we start to process $a_4$. If an internal node depends on a larger number of pruned nodes, the effectiveness of pruning will be degraded to some extent. To fix this challenging problem, we will introduce \textit{Probability Density Function} PDF that can be used to approximately compute the probability of a node, the result of which can be used to update the lower bound and upper bound of its ancestor nodes further. The details are provided and discussed with algorithms later.

\section{Probabilistic Inverted Index}\label{sec:index}

In this section, we describe our \textit{P}robabilistic \textit{I}nverted (PI) index structure for efficiently evaluating
PrTKQ queries over probabilistic XML data. 
In keyword search on certain XML data, inverted indexes are popular structures, e.g., \cite{DBLP:conf/sigmod/GuoSBS03,DBLP:conf/sigmod/XuP05}. The basic technique
is to maintain a list of lists, where each element in the outer list corresponds to a domain element (i.e., a keyword). Each inner list stores
the ids of XML nodes in which the given keyword occurs, and for each node, the frequencies or the weight at which the keyword appears or takes. 
In this work, we introduce a probabilistic version of this structure, in which we store for each keyword a list of node-ids. Along with each node-id, we store the \textit{probability values} that the subtree rooted at the node may contain the given keyword. 
The probability values in inner lists can be used to compute lower bound and upper bound on-the-fly during PrTKQ evaluation.

\begin{figure*}[htbp]
\centering
\includegraphics[scale=0.55]{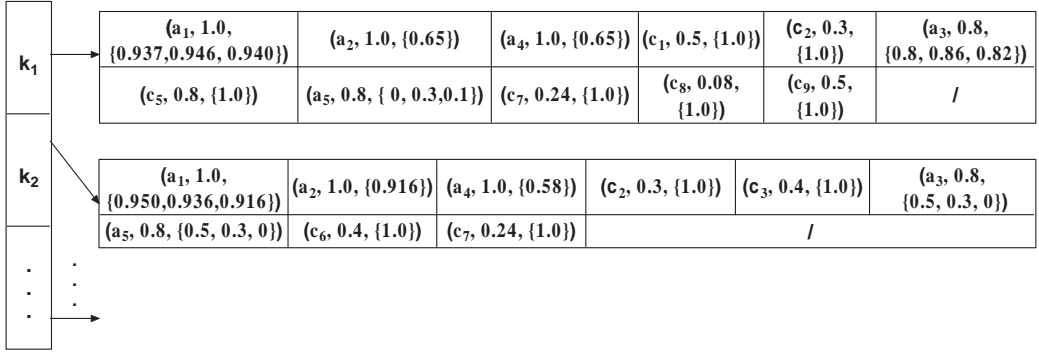}
\caption{A probabilistic Inverted Index}
\label{fig:invertindex}
\end{figure*}

Figure~\ref{fig:invertindex} shows an example of a probabilistic inverted index of the data in Figure~\ref{fig:pxml}. At the base of the structure is a list of keywords storing pointers to lists, corresponding to each term in the XML data $T$. This is an inverted array storing, for each term in $T$, a pointer to a list of triple tuples. In the list $k_i.list$ corresponding $k_i \in T$, the triple 
\scalebox{0.8}{
  \begin{tabular}{|c|}
    \hline
    ($v\_id$, Pr(path$_{r\rightarrow v}$), \{$p_1$, ...\})\\
    \hline
   \end{tabular}
 }
records the node $v$ \footnote{The symbol $v$ is used to represent a node's name or a node's id without confusions in the following sections. Here, $v$ is the id of the node $v$}, the conditional probability from the root to $v$, and the  probability set that may contain \textit{single probability value} or \textit{multiple probability value}. Single probability value represents that all the keyword instances in the subtree can be considered as independent in probability, e.g., the confidence of $a_2$ containing $k_1$ is \{0.65\}, while multple probability value means that the keyword instances belonging to different sets occur mutually, e.g., the confidence of $a_3$ containing $k_1$ is a set \{0.8, 0.86, 0.82\}, that represents the different possibilities of $k_1$ occurring in $a_3$.

\subsection{Basic Operations of Building PI Index}\label{subsec:operation}


To build PI index, we need to traverse the given XML data tree once in a bottom-up method. During the data traversal, we will apply the following operations that may be used solely or in their combinations. The binary operation X $\triangleright\triangleleft^{/}$ Y promotes the probability of Y to its parent node X. The binary operation X $\triangleright\triangleleft^{sibling}$ Y promotes the probabilities of two sibling nodes X and Y to their parent node. The n-ary case can be processed by calling for the corresponding binary cases one by one.  

Assume $v_1$ contains the keywords \{$k_1$, $k_2$, ..., $k_i$, ..., $k_m$\} and the conditional probability $Pr(path_{v_p->v_1})$ is $\lambda_1$; and 
$v_2$ contains the keywords  $\{k_i, k_{i+1}, ..., k_{m_i}\}$  and the conditional probability $Pr(path_{v_p->v_2})$ is $\lambda_2$.

\textbf{Operator1-}\textbf{v$_1$} $\Join^{sibling,IND}$ \textbf{v$_2$}:
If $v_1$ and $v_2$ are independent sibling nodes, we can directly promote their probabilities to their parent $v_p$, then we have,

\begin{equation}
Pr(k_j, v_p) = 
 \left\{ \begin{array}{ll}
							\lambda_1*Pr(k_j, v_1) &  j<i; \\
  1 - (1 - \lambda_1*Pr(k_j, v_1)) &\\
   *(1 - \lambda_2*Pr(k_j, v_2)) & i\leq j \leq m \leq m_i; \\
		\lambda_2 * Pr(k_j, v_2) & m \leq j \leq m_i; 
	\end{array}\right.																		
\end{equation}

\textbf{Operator2-}\textbf{v$_1$} $\Join^{/,IND}$ \textbf{v$_2$}: If $v_2$ is an independent child of $v_1$, we can directly promote the probability of $v_2$ to $v_1$, then we have, 

\begin{equation}
Pr(k_j, v_1) = 
 \left\{ \begin{array}{ll}
							Pr(k_j, v_1) &  j<i; \\
  1 - (1 - Pr(k_j, v_1)) &\\
   *(1 - \lambda_2*Pr(k_j, v_2)) & i\leq j \leq m \leq m_i; \\
		\lambda_2 * Pr(k_j, v_2) & m \leq j \leq m_i; 
	\end{array}\right.																		
\end{equation}

 
\begin{example}
Let's show the procedure of computing $c_1$$\Join^{sibling,IND}$$c_2$$\Join^{sibling,IND}$$c_3$ in Figure~\ref{fig:smallpxml} using Operator1 and Operator2. 
Firstly, we compute $c_1$$\Join^{sibling,IND}$$c_2$ and promote the probability of keywords to their parent $a_4$ by Operator1, i.e., $Pr(k_1, a_4)$ = 1 - (1 - 0.5*1.0)*(1 - 0.3*1.0) = 0.65 and $Pr(k_2, a_4)$ = 0.3. And then, we compute $a_4\Join^{/,IND}c_3$ using operator2, i.e., $Pr(k_2, a_4)$ = 1 - (1 - 0.3)*(1 - 0.4) = 0.58 while $Pr(k_1, a_4)$ do not change because $c_3$ only contains $k_2$ here.
And the conditional probability from the root to $a_4$ is 1.0. Therefore, $k_1 \rightarrow (a_4, 1.0, {0.65})$ and $k_2 \rightarrow (a_4, 1.0, {0.58})$ will be inserted in PI index, respectively.
\end{example}

 \textbf{Operator3-}\textbf{v$_1$} $\Join^{sibling,MUX}$ \textbf{v$_2$}: If $v_1$ and $v_2$ are two mutually-exclusive sibling nodes and $v_p$ is their parent, then we generate two parts in $v_p$ by $v_p\Join^{/,IND}v_1$ and $v_p\Join^{/,IND}v_2$, respectively.
 
 \textbf{Operator4-}\textbf{v$_1$} $\Join^{/,MUX}$ \textbf{v$_2$}: If $v_2$ is a mutually-exclusive child node of $v_1$, then we can get the aggregated probability by $v_1\Join^{/,IND}v_2$.

In the above four basic operators, we assume the terms independently appear in $v_1$ and $v_2$. When the nodes $v_1$ and $v_2$ contain mutually-exclusive parts, we need to deal with each part using the four basic operators.    
 
Given two independent sibling nodes $v_1$ ($\lambda_1$) and $v_2$ ($\lambda_2$) where only $v_2$ contains a set of mutually-exclusive parts $\{p_{m_1}, p_{m_2}, ...\}$ with conditional probability $\lambda_{m_i}$. In this case, we can apply the operation $v_1$ $\Join^{sibling,IND}$ $p_{m_i}$ for each part $p_{m_i}$. The computed results are maintained in different parts in their parent $v_p$.

\begin{example}
Consider an independent node $c_5$ and a node $a_5$ consisting of $c_6$, $c_7$ and $c_8$ in Figure~\ref{fig:pxml}. 
We first promote $c_6$, $c_7$ and $c_8$ to $a_5$ that consists of three parts: 
\scalebox{0.6}{
\begin{tabular}{|c|c|c|c|}
\hline
1 & \multicolumn{2}{|c|}{2} &3\\
\hline
$k_2$ & $k_1$ & $k_2$ & $k_1$ \\
\hline
0.5 & 0.3 & 0.3 & 0.1 \\
\hline
\end{tabular}}, as shown in Figure~\ref{fig:bound} - $a_5$. 
Because $c_5$ and $a_5$ are independent sibling nodes,
the operation $c_5 \Join^{sibling,IND} a_5$ can be called to compute the probability with regards to their parent $a_3$. To do this, we apply $c_5 \Join^{sibling,IND}$ part$_i$ for each part$_i$ $\in a_5$ using Operator1.
The results are shown in Figure~\ref{fig:bound} - $a_3$.  
After that, we can insert 
$k_1 \rightarrow (a_3, 0.8, {0.8, 0.86, 0.82})$ and $k_2 \rightarrow (a_3, 0.8, {0.5, 0.3, 0})$ into PI index.
\end{example}

If both $v_1$ and $v_2$ contain a set of mutually-exclusive parts, respectively, then we can do pairwise aggregations across the two sets of parts.
Building PI index needs to scan the given probabilistic XML data only once. Assume that the probabilistic XML has been encoded using probabilistic Dewey codes. The basic idea of building PI index is to progressively process the document nodes sorted by Dewey codes in ascending order, i.e., the data can be loaded and processed in a streaming strategy. 
When a leaf node $v_l$ is coming, we will compute the probability of each term in the leaf node $v_l$. After that, the terms with their probabilities in $v_l$ will be written into PI index. Next, we need promote the terms and their probabilities of $v_l$ to the parent $v_p$ of $v_l$ based on the operation types in Section~\ref{subsec:operation}. After the node stream is scanned completely, the building algorithm of PI index will be terminated. 
We don't provide the detailed building algorithm in this paper.



\subsection{Pruning Techniques using PI Index}
In this subsection, we first show how to prune the unqualified nodes using the proposed lower/upper bounds. And then, we explain how to compute lower/upper bounds, and how to update the upper/lower bounds based on intermediate results during the query evaluation.



By default, the node lists in PI index are sorted in the document order. $Pr(k_i, v)$ represents the overall probability of a keyword $k_i$ in a node $v$.
It is obvious that the overall probability of a keyword appearing in a node is larger than or equal to that of the keyword appearing in its descendant nodes. And the overall probability value for each keyword in a node can be computed and stored in PI index offline. 

Consider a node $v$ and a PrTKQ q containing a set of keywords $\{k_1, k_2, ..., k_t\}$. If all the terms in $v$ are independent, then we have, 

\begin{equation}\label{equ:indjoin}
LB(q,v) = \prod^{t}_{i=1}Pr(k_i, v)
\end{equation}

\begin{equation}
UB(q,v) = min\{Pr(k_i, v) | 1\leq i \leq t\}
\end{equation}

Most of the time, $v$ consists of a set of parts $\{vp_1, vp_2, ..., vp_m\}$ that are mutually-exclusive. In this case, the lower bound of $v$ would be generated from a part $vp_j$ that gives the highest lower bound value while the upper bound of $v$ would be generated from another part $vp_i$ that gives the highest upper bound value, in which $j$ may be equal to or not equal to $i$. 

\begin{equation}
LB(q,v) = max_{1\leq j \leq m}\{\prod^{t}_{i=1}Pr(k_i, vp_j)\}
\end{equation}
\begin{equation} 
UB(q,v) = max_{1\leq j \leq m}\{min\{Pr(k_i, vp_j) | 1\leq i \leq t\}\}
\end{equation}
Where $vp_j$ must satisfy the criteria: (1) $LB(q,v)>0$; (2) cannot find another part $vp_j'$ having $\prod^{t}_{i=1}Pr(k_i, vp_j') >0$ and $min\{Pr(k_i, vp_j') | 1\leq i \leq t\} > min\{Pr(k_i, vp_j) | 1\leq i \leq t\}$. Otherwise, UB($q,v$) and LB($q,v$) will be set as zero.

\begin{example}\label{exam:lubound}
Let's consider $a_3$ in Figure~\ref{fig:bound} as an example. The first and second parts can generate lower and upper bounds: Part 1 $\rightarrow$ LB($\{k_1,k_2\}$,$a_3$)=0.32, UB($\{k_1,k_2\}$,$a_3$)=0.4; and Part 2 $\rightarrow$ LB($\{k_1,k_2\}$,$a_3$)=0.206, UB($\{k_1,k_2\}$,$a_3$)=0.24. Because Part 1 can produce a higher upper bound than Part 2, the lower and upper bounds of $a_3$ will come from Part 1, which guarantees that $a_3$ can be a quasi-SLCA candidate with a higher probability. Since Part 3 does not contain full keywords, i.e., missing $k_2$, it cannot generate lower and upper bounds.
\end{example}   

\begin{property}
\textbf{[Upper Bound Usage]} A node $v$ can be filtered if the overall probability $Pr(k_i, v)$ of any keyword $k_i$ ( $k_i\in v$ and $k_i \in q$) is lower than the given threshold value $\sigma$, i.e., $\exists k_i$, $Pr(k_i, v) < \sigma$.
\end{property}
\begin{proof}
Since $Pr^{G}_{quasi-slca}(q,v) \leq min\{Pr(k_i, v)| k_i\in q\} \leq Pr(\forall k_i\in q, v)$, we have $min\{Pr(k_i, v)| k_i\in q\}$ as the upper bound probability of $v$ becoming a qualified quasi-SLCA node. Therefore, if a node $v$ holds the inequation $Pr(k_i, v) < \sigma$, then $Pr^{G}_{quasi-slca}(q,v)$ must be lower than $\sigma$. As such, $v$ can be filtered. 
\end{proof}

\begin{property}
\textbf{[Lower Bound Usage]} The nodes $v$ can be returned as required results if we have $LB(q, v) \geq \sigma$ and $UB(q, v_d) < \sigma$ where $v_d$ is any child or descendant node of $v$.
\end{property}
\begin{proof} $UB(q, v_d) < \sigma$ means that all the keyword nodes in the subtree rooted at $v$ will contribute their probabilities to node $v$. In other words, no decendant node of $v$ could be a quasi-SLCA so the lower bound probability $LB(q, v)$ will not be deducted. Therefore, if we have for $LB(q, v) \geq \sigma$, then $Pr^{G}_{quasi-slca}(q,v) \geq \sigma$. As such, $v$ can be returned as a quasi-SLCA result.
\end{proof}

\begin{example}
Let's continue Example~\ref{exam:lubound}. $a_3$ can be directly returned as a qualified answer for the given threshold $\sigma$( = 0.4). This is because $c_2$, $c_7$ and $a_5$ are filtered due to their upper bound less than the threshold $\sigma$( = 0.4). 
\end{example}

To update the lower/upper bound values during query evaluation, one way is to treat the different types of nodes differently, by which the updated lower/upper bounds may obtain better precision. But the disadvantage of this way is to easily affect the efficiency of bound update. This is because, given a current node having multiple quasi-SLCA nodes as its descendant nodes, it is required to know the detailed relationships (IND or MUX) among the multiple quasi-SLCA nodes. To avoid the disadvantage, we do not separate the different types of distributional nodes, under which the multiple quasi-SLCA nodes appear. In other words, we unify them into a uniform formula based on the following two  properties. 
 
\begin{property} \label{property:upperbound}
No matter node $v$ is an IND or ordinary or MUX node, we can update their upper bound values as follows:

\begin{equation}\label{equ:updateub}
UB^{'}(q,v) = UB(q,v) - 1 + \prod^{m}_{i=1}(1 - Pr^{G}_{quasi-slca}(q,v_{c_i})) 
\end{equation}
Where $Pr^{G}_{quasi-slca}(q,v_{c_i}) \geq \sigma$ should be held. 
\end{property}

\begin{proof}
According to the definition of upper bound, UB($q,v$) represents the maximal probability of $v$ being a quasi-SLCA node, which comes from the overall probability of a specific keyword. Therefore, the problem of updating upper bound can be alternatively considered as the percentage of the probability of the keyword has been used for the $v'$ descendant nodes becoming qualified quasi-SLCA nodes. If we know there are $m$ qualified descendant nodes of $v$ as returned answers, then we can compute their aggregated probabilities by $1 - \prod^{m}_{i=1}(1 - Pr^{G}_{quasi-slca}(q,v_{c_i}))$. Therefore, the upper bound can be updated as $UB(q,v) - 1 + \prod^{m}_{i=1}(1 - Pr^{G}_{quasi-slca}(q,v_{c_i}))$.    

Does the above update equation hold for MUX node? To answer this question, we utilize the properties in \cite{DBLP:conf/icde/LiLZW11}, from which we can compute the aggregated probability by using $\sum^{m}_{i=1}Pr^{G}_{quasi-slca}(q,v_{c_i})$. Therefore, we have 
$UB^{'}(q,v) = UB(q,v) - \sum^{m}_{i=1}Pr^{G}_{quasi-slca}(q,v_{c_i})$. The equation can be converted into $UB(q,v) - 1 + [1 - \sum^{m}_{i=1}Pr^{G}_{quasi-slca}(q,v_{c_i})]$.

Since $\prod^{m}_{i=1}(1 - Pr^{G}_{quasi-slca}(q,v_{c_i}))$ can be expressed as $1 - \sum^{m}_{i=1}Pr^{G}_{quasi-slca}(q,v_{c_i}) + \Delta$ where $\Delta$ is a positive value, i.e., $\geq$ 0, we can derive that $1 - \sum^{m}_{i=1}Pr^{G}_{quasi-slca}(q,v_{c_i})$ $\leq$ $\prod^{m}_{i=1}(1 - Pr^{G}_{quasi-slca}(q,v_{c_i}))$. As a consequence, we can obtain that $UB^{'}(q,v) = UB(q,v) - \sum^{m}_{i=1}Pr^{G}_{quasi-slca}(q,v_{c_i})$ = $UB(q,v) - 1 + [1 - \sum^{m}_{i=1}Pr^{G}_{quasi-slca}(q,v_{c_i})]$ $\leq$ $UB(q,v) - 1 + \prod^{m}_{i=1}(1 - Pr^{G}_{quasi-slca}(q,v_{c_i}))$. 

Therefore, $UB^{'}(q,v)$ = $UB(q,v)$ - 1 + $\prod^{m}_{i=1}$(1 - $Pr^{G}_{quasi-slca}$ ($q$, $v_{c_i}$)) holds for IND, ordinary and MUX nodes.
\end{proof}



\begin{property} 
No matter node $v$ is an IND or ordinary or MUX node, we can update their lower bound values as follows:

\begin{equation}\label{equ:updatelb}
LB^{'}(q,v) = LB(q,v) - \sum^{m}_{i=1}Pr^{G}_{quasi-slca}(q,v_{c_i}) 
\end{equation}
Where $Pr^{G}_{quasi-slca}(q,v_{c_i}) \geq \sigma$ should be held. 
\end{property}

\begin{proof}
For the lower bound update, we need to deduct the confirmed probability $[1 - \prod^{m}_{i=1}(1 - Pr^{G}_{quasi-slca}(q,v_{c_i}))]$ for IND nodes or $\sum^{m}_{i=1}Pr^{G}_{quasi-slca}(q,v_{c_i})$ for MUX nodes, from the original lower bound $LB(q,v)$. 
According to the procedure of the above proof, we have $\prod^{m}_{i=1}(1 - Pr^{G}_{quasi-slca}(q,v_{c_i}))$ $\geq$ $1 - \sum^{m}_{i=1}Pr^{G}_{quasi-slca}(q,v_{c_i})$. Consequently, we have the inequation, $1 - \prod^{m}_{i=1}(1 - Pr^{G}_{quasi-slca}(q,v_{c_i}))$ 
$\leq$ $1 - (1 - \sum^{m}_{i=1}Pr^{G}_{quasi-slca}(q,v_{c_i}))$ = $\sum^{m}_{i=1}Pr^{G}_{quasi-slca}(q,v_{c_i})$. Therefore, it is safe to use the right side to update the lower bound values.
\end{proof}

\begin{example}
Consider $a_4$ that has been computed and its probability is 0.44. Given threshold $\sigma$ (=0.4), $a_4$ is returned as a quasi-SLCA result. Consequently, we can update the lower/upper bound values of its ancestor $a_2$, i.e., $UB'$($\{k_1,k_2\}$, $a_2$) = 0.65 - 1 + (1 - 0.44) = 0.21 and $LB'(\{k_1,k_2\}, a_2)$ = 0.595 - 0.44 = 0.155. Since $UB'(\{k_1,k_2\},a_2) < \sigma$, $a_2$ can be filtered out effectively without computation. 
\end{example}

Property 3 is used to filter the unqualified nodes by reducing the upper bound value while Property 4 is used to quickly find the qualified required results by comparing the reduced lower bound value (for the probability of the remaining quasi-SLCAs) with the threshold value.

Sometimes, we need to calculate the probability distributions of keywords in a node if the given threshold $\sigma$ is in the range $(LB(q,v), UB(q,v)]$. The basic computational procedure is similar to the PrStack algorithm in \cite{DBLP:conf/icde/LiLZW11}. Different from the PrStack algorithm, we will introduce probability density function (PDF) to approximately calculate the probability for a node if the node depends on a large number of pruned descendent nodes. 
To decide when to invoke the PDF while avoiding the risk of reducing precision significantly, we would like to select and compute some descendant nodes that may contribute large probabilities to the node $v$. For the remaining descendant nodes, we may choose to invoke the PDF, by which we can reduce the time cost while still guarantee the precision to some extent. The detailed procedure will be introduced in the next section. 






\section{Prune-Based Probabilistic Threshold Keyword Query Algorithm}\label{sec:algorithms}

A key challenge of answering a PrTKQ is to identify the qualified result candidates and filter the unqualified ones as soon as possible. In this work, we address this challenge with the help of our proposed probabilistic inverted (PI) index. Two efficient algorithms are proposed, a comparable Baseline Algorithm and a PI-based Algorithm. 

\subsection{Baseline Algorithm}\label{subsec:basicalgorithm}
In keyword search on certain XML data, it is popular to use keyword inverted index retrieving the relevant keyword nodes, by which the keyword search results are generated based on different algorithms, e.g., \cite{DBLP:conf/sigmod/XuP05,DBLP:conf/sigmod/GuoSBS03,DBLP:conf/icde/BaoLCL09,DBLP:conf/edbt/LiLZW10}. In probabilistic XML data, \cite{DBLP:conf/icde/LiLZW11} proposed PrStack Algorithm to compute top-$k$ SLCA nodes. In this section, we propose an effective Baseline Algorithm that is similar the idea of PrStack Algorithm. 
To answer PrTKQ, we need to scan all the keyword inverted lists once. Firstly, the keyword-matched nodes will be read one by one based on their document order. After one node is processed, we check if its probability can be up to the given threshold value $\sigma$. If it is true, the node can be output as a quasi-SLCA node and its remaining keyword distributions (i.e., containing partial query keywords) can be continuously promoted to its parent node. Otherwise, we promote its complete keyword distributions (i.e., containing both all keywords or partial keywords) to its parent node. After that, the node at the top of the stack will be popped. Similarly, the above procedures will be repeated until all nodes are processed. The basic algorithm can be terminated when all nodes are processed. The detailed procedure is shown in Algorithm~\ref{algo:baseline}.

 
\begin{algorithm}[t]
  \small
  \caption{Baseline Algorithm } \label{algo:baseline} 
  \textbf{input:} a query $q=\{k_1, k_2, ..., k_m\}$ with threshold $\sigma$, keyword inverted (KI) index\\
  \textbf{output:} a set $R$ of quasi-SLCA nodes 
  
  \begin{algorithmic}[1]
			\STATE load keyword node lists $L$ = $\{l_1, l_2, ..., l_m\}$ from KI index;
			
			\STATE get the smallest Dewey $v$ from $L$;
			
			\STATE initiate a stack $S_1$ using $v$;
			
			\WHILE{$L \neq \phi$}
					\STATE get the next smallest Dewey $v$ from $L$;
						
					\WHILE{($S_1.$top() $\prec^{not}$ $v$)}
							
							\STATE $x$ = $S_1.$pop();
							\IF{$x$ contains full keywords and $Pr^{G}_{quasi-slca}(x) \geq \sigma$}
									\STATE output $x$ into $R$;
							\ENDIF
							
							\STATE promote the rest keyword distributions of $x$ to its parent $x_p$ using CombineProb($x$, $x_p$);
							
					\ENDWHILE
						 \STATE $S_1.$push($v$);
				\ENDWHILE

			\WHILE{$S_1 \neq \phi$}
					\STATE a new node $v \leftarrow$ $S_1.$pop();
					\IF{$v$ contains full keywords and $Pr^{G}_{quasi-slca}(v) \geq \sigma$}
									\STATE output $v$ into $R$;
							\ENDIF
							
							\STATE promote the rest keyword distributions of $v$ to its parent $v_p$ using CombineProb($v$, $v_p$);
					
			\ENDWHILE
  \RETURN $R$;

  \end{algorithmic}
  
\end{algorithm} 

Because Baseline Algorithm only needs to scan the keyword node lists once, it is a fast and simple algorithm. However, its core computation - keyword distribution computation would consume lots of time, which motivates us to propose the PI-based Algorithm that can quickly identify the qualified or unqualified candidates using offline computed PI index and only compute keyword distributions for a few candidates. Here, Baseline Algorithm is taken as a comparable base to show the pruning performance of the PI-based Algorithm described below.

\subsection{PI-based Algorithm}\label{subsec:online}

To efficiently answer PrTKQ, the basic idea of PI-based Algorithm is to read the nodes from keyword node lists one by one in a bottom-up strategy. 
For each node, we quickly compute its lower bound and upper bound by accessing PI index, which is far faster than computing the keyword distributions of the node directly. After comparing its lower/upper bounds with the given threshold value, we can decide if the node should be output as a qualified answer, skipped as an unqualified result, or cached as a potential result candidate. 
For example, if the current node's lower bound is larger than or equal to the threshold value, then the node can be output directly without further computation. This is because all its descendants have been checked according to the bottom-up strategy. If its upper bound is lower than the threshold value, then the node can be filtered out. Otherwise, it will be temporarily cached for further checking. 
Based on different cases, different operations would be applied. Only the nodes identified as potential result candidates need to be computed. Compared with Baseline Algorithm, PI-based algorithm can be accelerated significantly because Baseline Algorithm has to compute the keyword distributions for all nodes. 
The detailed procedure has been shown in Algorithm~\ref{alg:pibased}.

\begin{algorithm}[t]
  \small
  \caption{PI-based Algorithm } \label{alg:pibased} 
  \textbf{input:} a query $q=\{k_1, k_2, ..., k_m\}$ with threshold $\sigma$, keyword inverted (KI) index, PI index\\
  \textbf{output:} a set $R$ of quasi-SLCA nodes 
  
  \begin{algorithmic}[1]
			\STATE load keyword node lists $L$ = $\{l_1, l_2, ..., l_m\}$ from KI index; \label{line:maininitiate1}
			\STATE load probability node lists $PIL$ = $\{PIL_1, PIL_2, ..., PIL_m\}$;
			
			\STATE get the smallest Dewey $v$ from $L$;
			
			\STATE initiate a stack $S_1$ using $v$ and an empty stack $S_2$; \label{line:maininitiate2}
			
			\WHILE{$L \neq \phi$}
								 
						
							\STATE get the next smallest Dewey $v$ from $L$ again;
						
							\WHILE{($S_1.$top() $\prec^{not}$ $v$)}
							
									\STATE $x$ = $S_1.$pop();
									\STATE UB($q$,$x$) and LB($q$,$x$) $\leftarrow$ ComputeBound($x$, $\{PIL_i(x)\}$);
									
									\IF{LB($q$,$x$)$\geq \sigma$} \label{line:lbdasigma1}
											\STATE output $x$ into $R$;
											
											\STATE UpdateBound($\{v_a\in S_1|v_a \prec x\}$, LB($q$,$x$), UB($q$,$x$));\label{line:lblargesigma}
													
											\STATE $S_2.$pop($v_d \in S_2|v_d \succ x$);										
									
									\ELSIF{UB($q$,$x$) $\geq \sigma$ $>$ LB($q$,$x$)}
									
											\STATE $Prob(x)$ $\leftarrow$ ComputeProbDist($x$, $S_2$); 
											\IF{$Prob(x) \geq \sigma$}
											
													\STATE output $x$ into $R$;
											
													\STATE UpdateBound($\{v_a\in S_1|v_a \prec x\}$, $Prob(x)$);\label{line:lblesssigma}
															
													\STATE $S_2.$pop($v_d \in S_2|v_d \succ x$);
											
											\ENDIF
									\ELSE 
									
													\STATE $S_2.push$($x$);
																
									\ENDIF \label{line:lbdasigma2}
						 \ENDWHILE
						 \STATE $S_1.$push($v$);
				\ENDWHILE

			\WHILE{$S_1 \neq \phi$}
					\STATE a new node $v \leftarrow$ $S_1.$pop();
					\STATE UB($q$,$v$) and LB($q$,$v$) $\leftarrow$ ComputeBound($v$, $\{PIL_i(v)\}$);
					
					\STATE process the node $v$ using the same codes in Line~\ref{line:lbdasigma1} - Line~\ref{line:lbdasigma2};
					
			\ENDWHILE
  \RETURN $R$;

  \end{algorithmic}
  
\end{algorithm} 

\subsubsection{Detailed Procedure of PI-based Algorithm}
In Algorithm~\ref{alg:pibased},  Line~\ref{line:maininitiate1}-Line~\ref{line:maininitiate2} show that the procedures of initiating PI-based Algorithm. We first load the keyword node lists $L$ from KI index and probability node lists $PIL$ from PI index. And then we take the smallest node $v$ from $L$ to initiate a stack $S_1$ that is set using the dewey codes of $v$. Another stack $S_2$ is used to maintain the temporary filtered nodes. After that, the PI-based Algorithm is ready to start. 

Next, we need to check each node in $L$ in document order.
Different from Baseline Algorithm, we only compute the keyword distribution probabilities for a few nodes that are first identified using the lower bound and upper bound in $PIL$. 
Consider $v$ be the next smallest node to be processed. We compare it with the node $x$ in stack $S_1$. If $v$ is the descendant node of $x$, then $v$ will be pushed into $S_1$ and get the next smallest node from $L$. Otherwise, we pop out $x$ from $S_1$ and check if it is a qualified quasi-SLCA answer. In Baseline Algorithm, it will compute the keyword distributions of $x$ and combine its remaining distributions and the distribution of its parent based on promotion operations. Different from Baseline Algorithm, PI-based Algorithm will quickly compute the upper bound UB($q$,$x$) and lower bound LB($q$,$x$) using $PIL$, which is used to differentiate the nodes as qualified nodes - output, unqualified nodes - filter and uncertain nodes - to be further checked. By doing this, only a few nodes need to be computed. Since bound computation is far faster than computation of keyword distribution, lots of run time cost can be saved in PI-based Algorithm.  
Line~\ref{line:lbdasigma1}-Line~\ref{line:lbdasigma2} show the detailed procedures. If the lower bound LB($q$,$x$) is larger than or equal to the given threshold value $\sigma$, then $x$ can be output as a qualified quasi-SLCA answer without computation. 
At this moment, the lower bound LB($q$,$x$) can be taken as the temporary probability of $x$ being a quasi-SLCA result because the exact probability of $x$ is delayed until we need to calculate its exact probability value. Subsequently, the temporary probability value LB($q$,$x$) and the probabilities of $x'$ descendant quasi-SLCA results can be used to update the lower/upper bounds of the ancestors of $x$ in stack $S_1$ based on Equation~\ref{equ:updatelb} and Equation~\ref{equ:updateub}, respectively.
If the lower bound LB($q$,$x$) is lower than $\sigma$ while the upper bound UB($q$,$x$) is larger than or equal to $\sigma$, then we need to compute the keyword distributions of $x$ using the cached descendant nodes in $S_2$. Based on the computed probability $Prob(x)$ of $x$, it can be decided to be output as a qualified answer or filtered as an unqualifed candidate. If the upper bound UB($q$,$x$) is lower than $\sigma$, then $x$ will be pushed into $S_2$ for the possible computation of its ancestors.  

There are two main functions in PI-based Algorithm. The first one is ComputeProbDist($v$, $S_2$) for computing the probability of full keyword distribution of $v$ using the descendant nodes in $S_2$. The second is UpdateBound($\{v_a \prec v|v_a\in S_1\}$, LB($q$,$v$) or Prob($q$,$v$)) for updating the bounds of the nodes to be processed.


\subsubsection{Function ComputeProbDist()}
The function ComputeProbDist($v$, $S_2$) can be implemented in two ways, \textit{Exact Computation} or \textit{Approximate Computation}. 

\textit{Exact Computation} is to actually calculate the probability of $v$ being a quasi-SLCA node by scanning all the nodes in the stack $S_2$ that maintains the descendant nodes of $v$. The processing strategy is similar to Baseline Algorithm in Section~\ref{subsec:basicalgorithm}. In other words, it needs to visit the nodes in $S_2$ one by one and compute the local keyword distribution of each node, and then promotes the intermediate results to its parent. After all nodes in $S_2$ are processed, the probability of $v$ will be obtained because it aggregates all the probabilities from its descendant nodes.

\textit{Approximate Computation} is to approximately calculate the probability of $v$ being a quasi-SLCA node based on a partial set of nodes in the stack $S_2$ that maintains the descendant nodes of $v$. The approximate computation can be made according to different distribution types, e.g., uniform distributions, piecewise polynomials, poisson distributions, etc. In this work, we consider normal or Gaussian distributions in more detail. 

As we know Gaussian distribution is considered the most prominent probability distribution in statistics. However, the PDF of Gaussian distribution cannot be applied to PrTKQ over probabilistic XML data directly due to two main challenges. The first challenge is to simulate the continuous distributions using discrete distributions based on the real conditions in order to reduce the approximate errors as much as possible, and the second is to embody the multiple keyword variables in the PDF.

Generally, the probability density function of a Gaussian distribution $N(\mu, \sigma^2)$ of mean $\mu$ and variance $\sigma^2$ is:

\begin{equation}\label{equ:gaussian}
f(x) = \frac{1}{\sqrt{2 \pi \sigma^2}}e^{-(x-\mu)^2/(2\sigma^2)}
\end{equation}

\textbf{Addressing Challenge 1:} The density function has a shape of a bell centered in the mean value $\mu$ with variance $\sigma^2$. Based on the definition of Gaussian distribution, the Gaussian distribution is often used to describe, at least approximately, measurements that tends to cluster around the mean. Therefore, consider the mean $\mu$ be the partial computed probability value of $v$ be a quasi-SLCA node, which guarantees the real probability value will not be significantly different from the probability base that has already been calculated based on promising descendant nodes. The value of the variance $\sigma^2$ can be chosen from the range [1-\#computed descendant nodes/\#total descendant nodes, 1] based on the visited/unvisited descendant nodes in $S_2$. This is because the more the descendant nodes are actually computed, the higher the percentage of the values would be drawn within one standard deviation $\sigma$ away from the mean. Extremely, if all descendant nodes are computed actually, 100\% of values can be drawn within one stardard deviation. Therefore, we select and compute a few descendant nodes of $v$ from $S_2$, which can contribute relatively higher probabilities to make $v$ a quasi-SLCA node. In this work, we use heuristic method to select a few descendant nodes with the higher probabilities of single keywords in the descendant nodes of $v$. And then, we take the partially computed probability as the base of the probability density function of a Gaussian distribution.  
 
 Consider $v$ be a node to be evaluated and UB($q$,$v$)~\footnote{Note that UB($q$,$v$) has been updated if $v$ has descendant nodes that are qualified answers, i.e., it minus the probability contributions of the qualified answers.} be its current upper bound value. We have,
 
 \begin{equation}\label{equ:prgaussian}
 Pr^{G, Gaussian}_{quasi-slca}(q,v) = \int^{UB(q,v)}_{0}f(x)dx 
 \end{equation}
 
 After substituting Equation~\ref{equ:gaussian} into Equation~\ref{equ:prgaussian}, we get,
 
 \begin{equation}\label{equ:ourprgaussian}
 Pr^{G, Gaussian}_{quasi-slca}(q,v) = \int^{UB(q,v)}_{0}\frac{1}{\sqrt{2 \pi \sigma^2}}e^{-\frac{(x-\mu)^2}{2\sigma^2}}dx
 \end{equation}
 Where $\mu$ is the partially computed probability, $\sigma^2$ is set as 1-\#computed descendant nodes/\#total descendant nodes.
 
 \textbf{Addressing Challenge 2:} To embody all the keyword variables in the PDF, we introduce the joint/conditional Gaussian distribution based on the work in \cite{books/mit/DbJt}. Assume a PrTKQ contains two keywords $k_x$ and $k_y$. We have the conditional PDF as follows.

 \begin{equation}\label{equ:condgaussian}
f_{Y|X}(y|x) = \frac{1}{\sqrt{2 \pi (1-\rho^2) \sigma_Y^2}}e^{-\frac{[(y-\mu_Y) - (\rho (\frac{\sigma_Y}{\sigma_X}) (x - \mu_X))]^2}{2\sigma_Y^2}}
\end{equation} 
 
Since $f(x, y) = f_X(x)*f_{Y|X}(y|x)$, after substituting Equation~\ref{equ:gaussian} into Equation~\ref{equ:condgaussian}, we get 
 
 \begin{equation}\label{equ:2vgaussian}
 \begin{array}{l}
f(x, y) = \\
\frac{1}{2 \pi \sigma_X \sigma_Y \sqrt{1-\rho^2}}e^{-\frac{(x - \mu_X)^2}{2\sigma_{X}^2}-\frac{[(y-\mu_Y) - (\rho (\frac{\sigma_Y}{\sigma_X}) (x - \mu_X))]^2}{2\sigma_Y^2}}
\end{array}
\end{equation} 

If we make an assumption that $x$ and $y$ are independent keyword variables i.e., $\rho = 0$, and assume $\mu_X = \mu_Y = \mu$ and $\sigma_X = \sigma_Y = \sigma$, then we have

 \begin{equation}\label{equ:2vindgaussian}
f(x, y) = \frac{1}{2 \pi \sigma^2}e^{-\frac{(x - \mu)^2+(y-\mu)^2}{2\sigma^2}}
\end{equation} 
 
 Therefore, Equation~\ref{equ:2vindgaussian} can be easily extended to multiple keyword variables that are assumed as independent. We can compute the probability of $v$ w.r.t. a PrTKQ $\{k_1, k_2, ..., k_t\}$.
  
 \begin{equation}\label{equ:ourfinalprgaussian}
\begin{array}{l}
Pr^{G, Gaussian}_{quasi-slca}(q,v) = \\ \int^{UB(q,v)}_{0}...\int^{UB(q,v)}_{0}\frac{e^{-\frac{(x_1-\mu)^2+...+(x_t-\mu)^2}{2\sigma^2}}}{(2\pi)^{t/2} \sigma^t}dx_1...dx_t
\end{array}
\end{equation}
 Where $\mu$ is the partially computed probability, $\sigma^2$ is set as 1-\#computed descendant nodes/\#total descendant nodes.

 In the experiments, we call Matlab from Java to calculate Equation~\ref{equ:ourfinalprgaussian}. The estimated results are used to show the comparison between the actual computation and approximation computation. The results verify the usability of Gaussian distribution to measure the probability.

\subsubsection{Function UpdateBound()} 
 For each ancestor node $v_a\in S_1$ ($v_a \prec v$), we need to update the upper bounds and lower bounds using Function UpdateBound() based on Equation~\ref{equ:updateub} and Equation~\ref{equ:updatelb}, respectively. 
 To guarantee the completeness of the answer set, the parameters of the function may be different based on conditions. For example,  
 if LB($q$,$v$) is larger than or equal to the threshold value $\sigma$ as shown in Algorithm~\ref{alg:pibased}: Line~\ref{line:lblargesigma}, 
 then the probability value LB($q$,$v$) is used to update the upper bounds of $v'$ ancestors while the probability value UB($q$,$v$) is used to update the lower bounds of $v'$ ancestors; 
if LB($q$,$v$) is smaller than $\sigma$ and UB($q$,$v$) is larger than or equal to $\sigma$ as shown in Algorithm~\ref{alg:pibased}: Line~\ref{line:lblesssigma}, the actual or approximate probability value $Prob(v)$ computed by Function ComputeProbDist($v$, $S_2$) will be utilized to update the upper/lower bounds of $v'$ ancestors together.  
 
 

Here, we use two hashmaps to implement Function UpdateBound(). For a node, one hashmap is used to cache the dewey of the node as a key, and the lower/upper bounds as a value where the bounds are computed based on PI index. Another hashmap is used to record the probability that the descendants of the node having been identified as qualified quasi-SLCA answers. When a node is coming, we can quickly get the updated lower/upper bounds based on the two hashmaps.

\section{Experimental Studies}\label{sec:experimentalresults}

We conduct extensive experiments to test the performance of our algorithms: Baseline Algorithm (BA); PI-based Exact-computation Algorithm (PIEA) that implements Function ComputeProbDist() by exactly computing the probability distributions of the keyword matched nodes; and PI-based Approximate-computation Algorithm (PIAA) that makes approximated computation based on the Gaussian distribution of keywords while still exactly computing the probability distributions of the keyword matched nodes that have the higher probabilities. All these algorithms were implemented in Java and run on a 3.0GHz Intel Pentium 4 machine with 2GB RAM running Windows 7.

\subsection{Dataset and Queries}

We use two real datasets, DBLP~\cite{dblp:dblp} and Mondial~\cite{mondial}, and a synthetic XML benchmark dataset XMark~\cite{xmark} for testing the proposed algorithms. For XMark, we also generate four datasets with different sizes. The three types of datasets are selected based on their features. DBLP is a relatively shallow dataset of large size; Modial is a deep and complex, but small dataset; XMark is a balanced dataset with varied depth, complex structure and varied size. Therefore, they are chosen as test datasets.

For each XML dataset used, we generate the corresponding probabilistic XML tree, using the same method as used in \cite{DBLP:conf/sigmod/KimelfeldKS08}. We visit the nodes in the original XML tree in pre-order way. We first set the random ratio of IND:MUX:Ordinary as 3:3:4. For each node $v$ visited, we randomly generate some distributional nodes with ``IND'' or ``MUX'' types as children of $v$. 
 Then, for the original children of $v$, we choose some of them as the children of the new generated distributional nodes and assign random probability distributions to these children with the restriction that the sum of them for a MUX node is no greater than 1. 
The generated datasets are described in Table~\ref{tab:dataset}. And  we select terms and construct a set of keyword queries to be tested for each dataset.
Due to the limited space, we only show six of these queries for each dataset. For each different sets of queries, the terms in the first two queries have small size of keyword matched nodes; the terms of the middle two queries relate to a medium size of keyword matched nodes; the terms of the last queries are based on the computation of a larger number of keyword matched nodes.

\begin{table}[t]
  \renewcommand{\arraystretch}{1.2}
  \small
  \centering
  \caption{Properties of PrXML data}
  \label{tab:dataset}
    \scalebox{0.6}{
  \begin{tabular}{||c|l|rrrr||}
    \hline
    ID & name & size & \#IND & \#MUX & \#Ordinary \\[0.5ex]
    \hline 
    Doc1 & XMark & 10M  & 26k   & 26k  &   145k      \\ 
    Doc2 &      & 20M  & 54k   & 52k  &   200k      \\
    Doc3 &      & 40M  & 98k  & 100k & 606k     \\ 
    Doc4 &      & 80M  & 329k  & 368k & 1M          \\ 
    \hline
    Doc5 &Modial &1.2M & 8k      & 9k    & 20k \\
    \hline
    Doc6 &DBLP & 136M & 759k &  589k &  3M\\
    \hline
     Doc7 &INEX & 5,898M & 13M &  10M &  52M\\
     \hline
  \end{tabular}}
\end{table}

\subsection{Varying Keyword Queries}

\begin{figure}[ht]
  \centering
  \subfigure[XMark]{\label{fig:querytimexmark} 
    \includegraphics[scale=0.6]{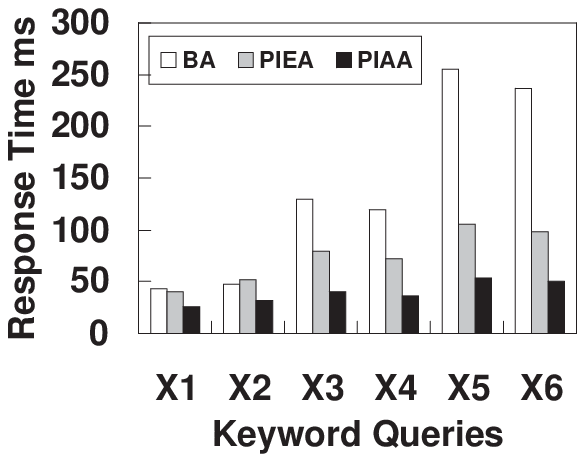}}
  \subfigure[Precision\&Recall]{\label{fig:queryprxmark}
    \includegraphics[scale=0.6]{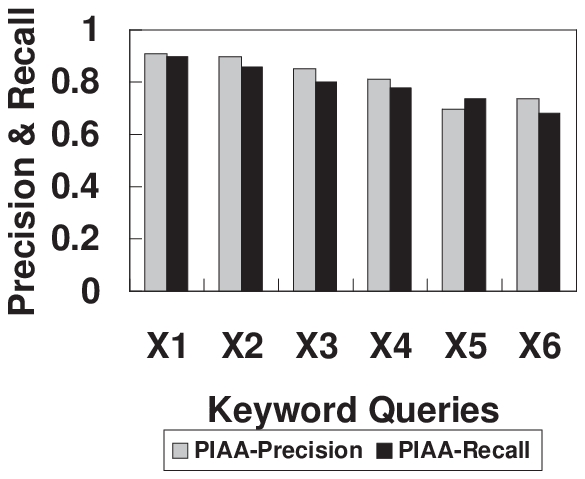}} 
    \\
  \subfigure[Mondial]{\label{fig:querytimemodial}
    \includegraphics[scale=0.6]{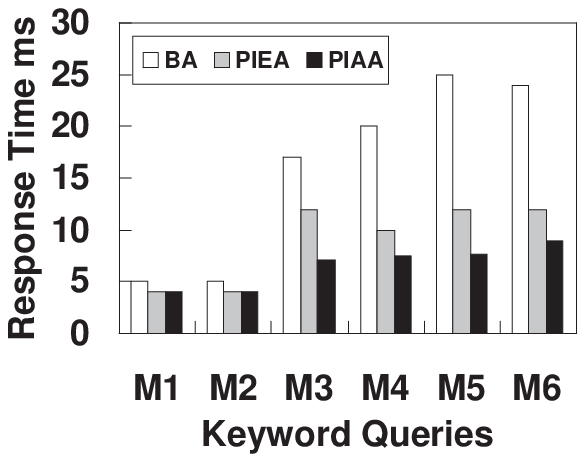}}  
  \subfigure[Precision\&Recall]{\label{fig:queryprmodial}
    \includegraphics[scale=0.6]{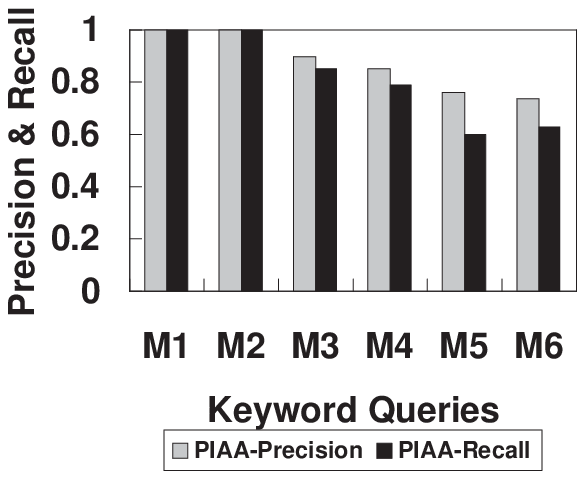}}     
      \\
  \subfigure[DBLP]{\label{fig:querytimedblp}
    \includegraphics[scale=0.6]{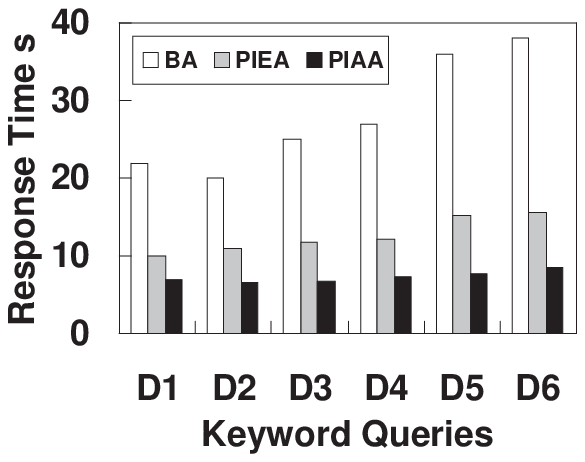}}  
  \subfigure[Precision\&Recall]{\label{fig:queryprdblp}
    \includegraphics[scale=0.6]{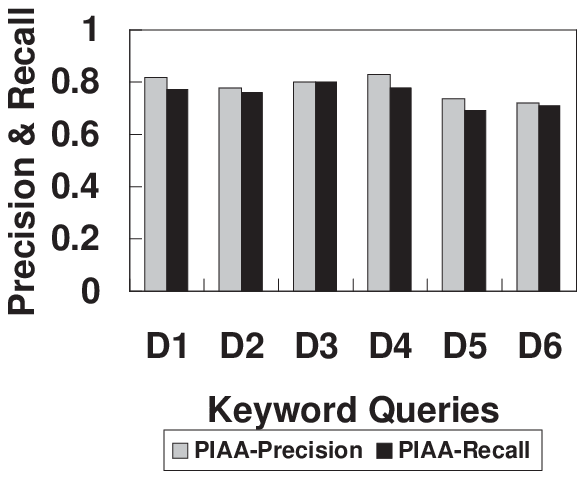}} 
          \\
  \subfigure[INEX]{\label{fig:querytimeinex}
    \includegraphics[scale=0.6]{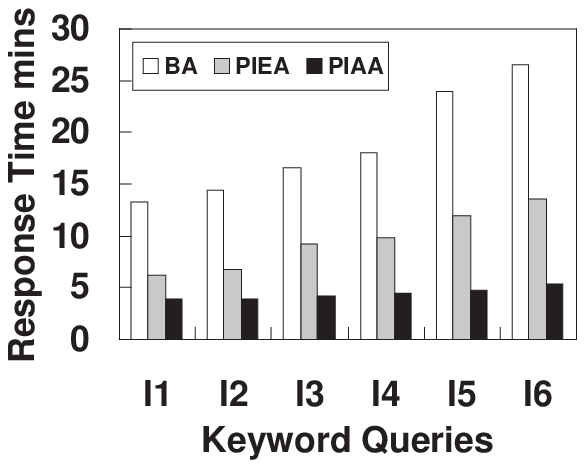}}  
  \subfigure[Precision\&Recall]{\label{fig:queryprinex}
    \includegraphics[scale=0.6]{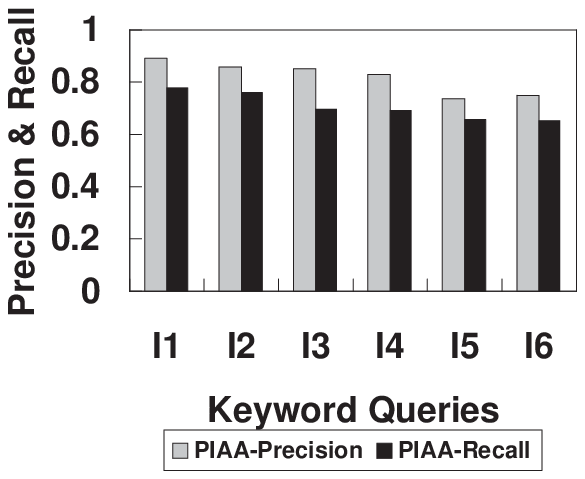}} 
     \caption{Evaluation of Keyword Queries over XMark20M, Mondial, DBLP, INEX where $\sigma$=0.3}
  \label{fig:varyquery} 
\end{figure}

 Figure~\ref{fig:varyquery} shows the experimental results when we run the 18 queries over the selected three datasets where $X$ represents the queries over 20MB XMark dataset, $M$ represents the queries over Mondial dataset, $D$ represents the queries over DBLP dataset and $I$ represents the queries over INEX dataset. And the required threshold value $\sigma$ is set as 0.3. From the results, we can find that compared with the BA algorithm, most of time the PIEA algorithm can reduce the response time by nearly 40\% using the pruning techniques based on the updated lower/upper bounds. The PIAA algorithm can further improve the time efficiency by about 20\% with the assumption of probability distribution of keywords. For $X1$, $X2$ and $M_1$, $M_2$, the response time of BA algorithm is approaching to the time cost of the other two algorithms. Especially for query $X2$, PIEA algorithm is overwhelmed by BA algorithm. This is because both the number of keyword-matched nodes and the size of answer sets are smaller than the other queries. From the four figures on the left side of Figure~\ref{fig:varyquery}, we find that the scalability of PIEA and PIAA algorithms is much better than that of BA algorithm by testing the queries with different sizes of answer sets.

 To measure the precision and recall of PIAA algorithm, we utilize the P\&R equations in information retrieval area as follows.
 
 $Precision = \frac{|R_{BA}\cap R_{PIAA}|}{|R_{PIAA}|}$; 
 $Recall = \frac{|R_{BA}\cap R_{PIAA}|}{|R_{BA}|}$
 
 Because PIEA algorithm can find the same results with BA algorithm by exactly computing the required probability distributions, Figure~\ref{fig:varyquery} only demonstrates the precision and recall of PIAA algorithm for different queries over each dataset. From the experimental results, we find that the precision and recall can reach up to at least 0.7 for XMark, 0.6 for Mondial, 0.7 for DBLP, and 0.66 for INEX, respectively. Sometimes, it can be up to 0.9 at most, e.g., X1, X2, M1, M2, I1, I2, etc. Comparing all the tested queries, we can get a general conclusion that the precision and recall will be decreased with the increase of potential result size. However, from the experiments, they will not be lower than 0.6 because (1) the results with higher probabilities are exactly selected and computed, which does not need to depend on the Gaussian assumption; (2) the rest minor results are estimated by using Gaussian assumption over the keyword distributions that have been excluded by the results with higher probabilities. In other words, PIAA strategy can return the percentage ($\geq$0.6) of significant results, but may underestimate the minor results.   
 

\subsection{Varying Threshold Values}

\begin{figure}[ht]
  \centering
  \subfigure[XMark]{\label{fig:thresholdtimexmark}
    \includegraphics[scale=0.6]{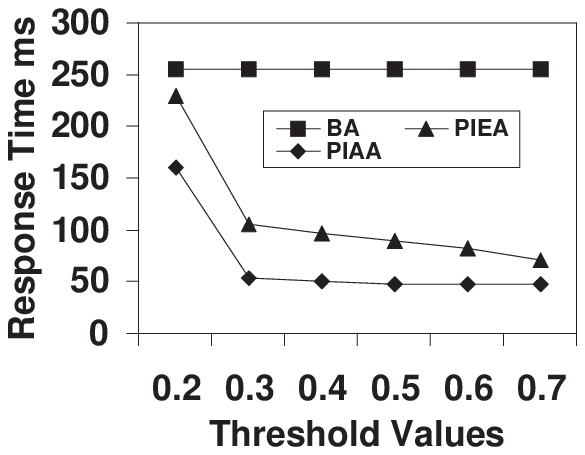}
    }
  \subfigure[Precision\&Recall]{\label{fig:thresholdprxmark}
    \includegraphics[scale=0.6]{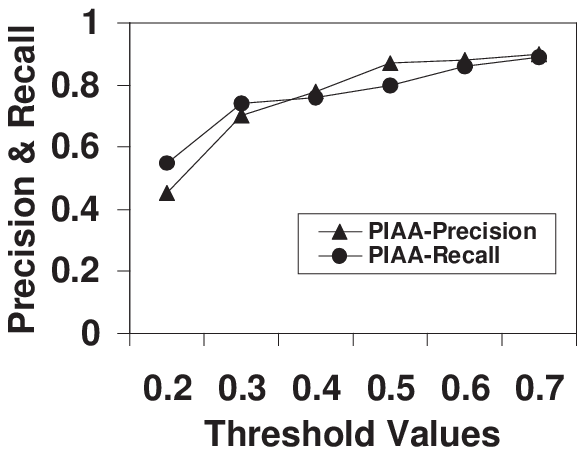}
    } 
    \\
     \subfigure[Mondial]{\label{fig:thresholdtimemodial}
    \includegraphics[scale=0.6]{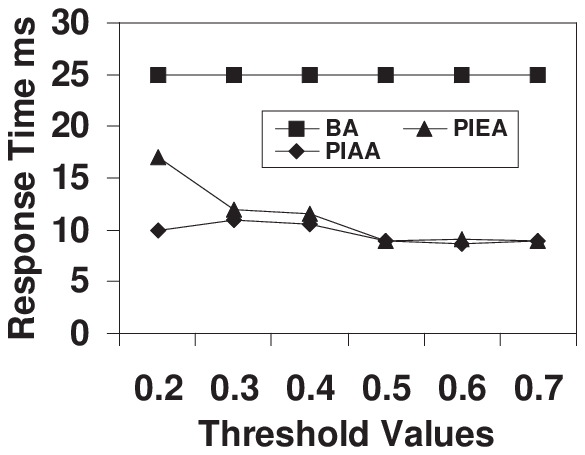}
    }
  \subfigure[Precision\&Recall]{\label{fig:thresholdmemorymodial}
    \includegraphics[scale=0.6]{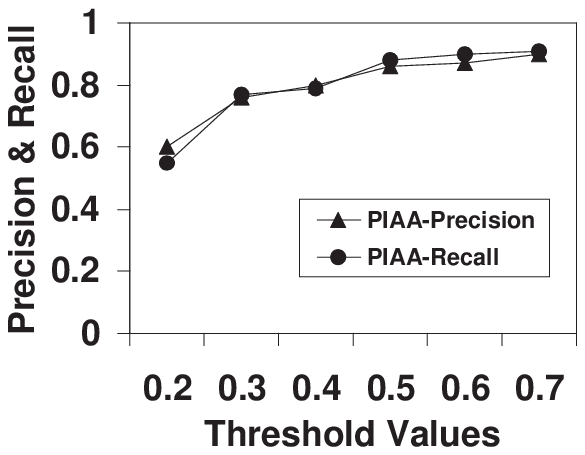}
    } 
    \\
     \subfigure[DBLP]{\label{fig:thresholdtimedblp}
    \includegraphics[scale=0.6]{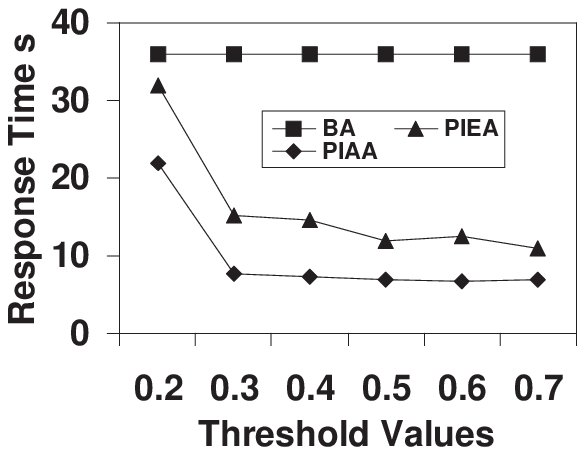}
    }
  \subfigure[Precision\&Recall]{\label{fig:thresholdmemorydblp}
    \includegraphics[scale=0.6]{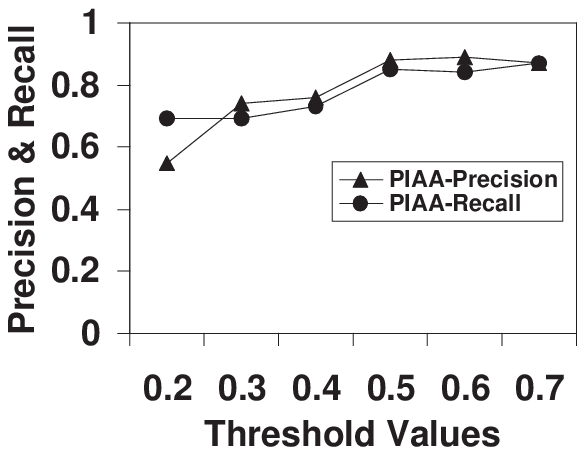}
    }  
       \\
     \subfigure[INEX]{\label{fig:thresholdtimeinex}
    \includegraphics[scale=0.6]{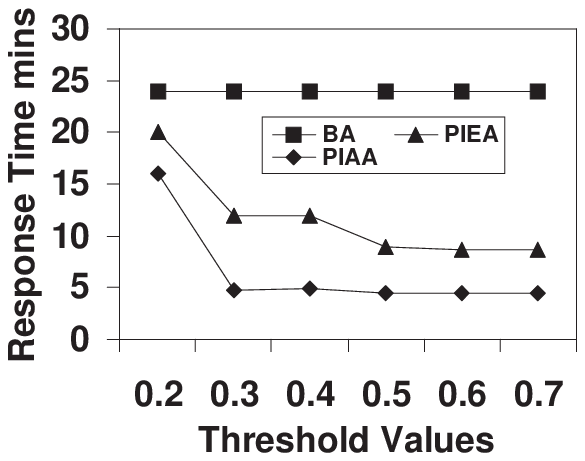}
    }
  \subfigure[Precision\&Recall]{\label{fig:thresholdmemoryinex}
    \includegraphics[scale=0.6]{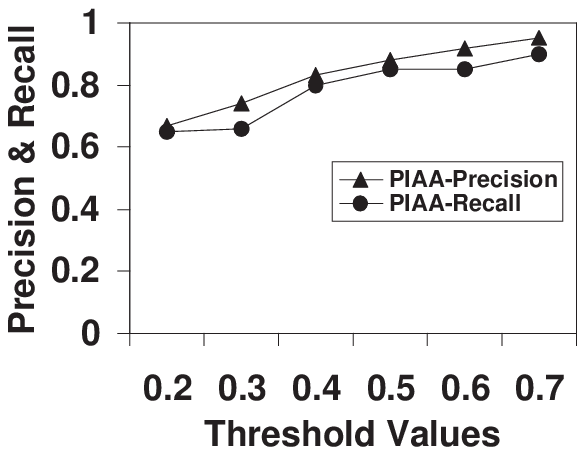}
    }   
     \caption{Time, Precision and Recall vs. Varied Threshold}
  \label{fig:threshold} 
\end{figure}

To test the adaptability of the proposed algorithms to threshold query, we test the changes of response time and precision\&recall with the increase of threshold value. 
Figure~\ref{fig:threshold} shows the experimental results when the threshold value varies from 0.2 to 0.7 for queries X5, M5, D5 and I5. The left four figures in Figure~\ref{fig:threshold} show that PIEA and PIAA algorithms can overwhelm BA algorithm greatly with the increase of threshold value. This is because BA algorithm has to scan and compute all the relevant nodes while PIEA and PIAA algorithms can skip more nodes when the threshold value becomes large. However, when the threshold value is up to 0.5, the change of the time cost will be smooth because once a quasi-SLCA node is identified, its ancestor nodes can be skipped definitely, which is true for the threshold values larger than 0.5. 
From the right four figures in Figure~\ref{fig:threshold}, we can find that the precision and recall will be affected by the change of threshold values. When the threshold value reaches up to 0.5, the precision and recall can be up to 0.8 at least. On the contrary, if the threshold value is lower than 0.2, the precision and recall would be decreased to 0.5 based on the selected datasets.

\subsection{Varying Probabilistic Document Size}

\begin{figure}[ht]
  \centering
  \subfigure[XMark (X3, $\sigma$=0.3)]{\label{fig:timedocsizexmarkq3} 
      \includegraphics[scale=0.6]{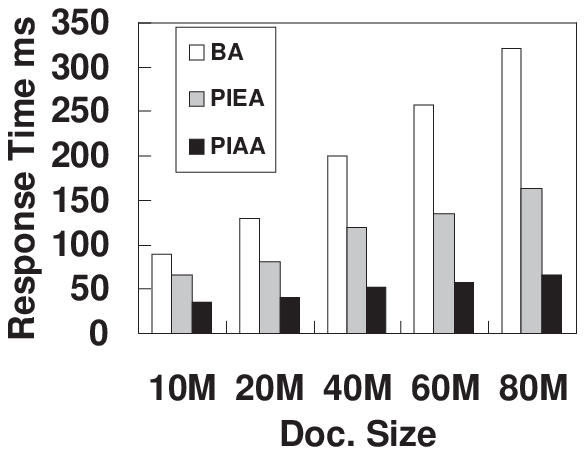}} 
     \subfigure[A Variant of F-Measure]{\label{fig:fmeasure}
     \includegraphics[scale=0.6]{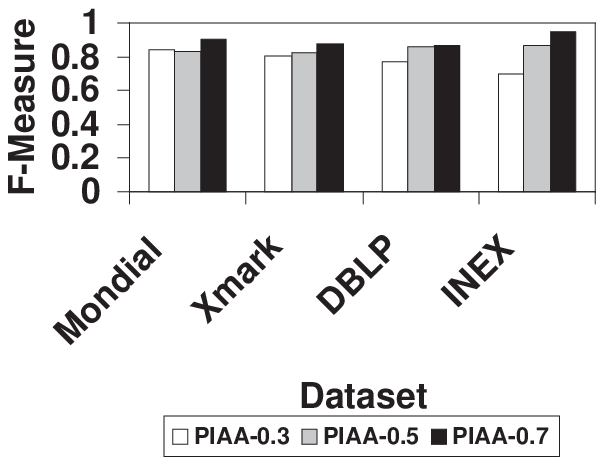}}
   \caption{Response Time and F-Measure for different datasets}   
  \label{fig:dataset}
\end{figure}

We firstly take XMark dataset as an example to test the performance of the three algorithms when we increase the document size. We test all the six queries of XMark dataset, but in this paper, we only show the results of the query $X3$ where the threshold value is specified as 0.3. From Figure~\ref{fig:timedocsizexmarkq3}, we can see that the response time of all the three algorithms will be increased when the document size increases from 10MB to 80MB. 
However, the increase of PIEA and PIAA algorithms is much slower. Particularly, PIAA just changes a bit. 
The comparison illustrates that PIEA and PIAA algorithms can obtain much better performance than BA algorithm. In addition, all algorithms show linear degradation, i.e., they have the similar scalability.

Secondly, we evaluate the precision and recall of PIAA algorithm using a variant of F-measure that aggregates the precision and recall of all queries together.

$F-measure = 2*\frac{\overline{P(q_i)}*\overline{R(q_i)}}{\overline{P(q_i)}+\overline{R(q_i)}}$

Where $\overline{P(q_i)} = \sum_{1}^{6}(P(q_i))/6$, and $\overline{R(q_i)} = \sum_{1}^{6}(R(q_i))/6$. 

To evaluate the F-measure of PIAA algorithm, we test 24 queries with different threshold values: 0.3, 0.5 and 0.7. From the results in Figure~\ref{fig:fmeasure}, we can find that the F-measure can be over 0.75 for all the four datasets.

\section{Related Work} \label{sec:relatedwork}

The topic of probabilistic XML has
been studied recently. Many models have been proposed, together
with structured query evaluations. Nierman et
al. \cite{DBLP:conf/vldb/NiermanJ02} first introduced a probabilistic XML model, ProTDB, with the probabilistic types IND - \textit{independent} and MUX - \textit{mutually-exclusive}. Hung et al. \cite{DBLP:conf/icde/HungGS03} modeled
the probabilistic XML as directed acyclic graphs, supporting arbitrary distributions over 
sets of children. Keulen et al. \cite{DBLP:conf/icde/KeulenKA05} used a probabilistic tree approach for data integration where its probability and possibility nodes are similar to MUX and IND, respectively. 
Cohen et al. \cite{DBLP:journals/tods/CohenKS09} incorporated a set of constraints to express more complex
dependencies among the probabilistic data. They also proposed efficient
algorithms to solve the constraint-satisfaction, query evaluation,
and sampling problem under a set of constraints. In \cite{DBLP:conf/sigmod/KimelfeldKS08},
Kimelfeld et al. summarized and extended the probabilistic XML models
previously proposed, the expressiveness and tractability of queries
on different models are discussed with the consideration of IND and MUX. \cite{DBLP:conf/vldb/KimelfeldS07} studied the problem of evaluating twig queries over probabilistic XML that may return incomplete or partial answers with respect to a probability threshold to users. \cite{DBLP:conf/edbt/ChangYQ09} proposed and addressed the problem of ranking top-k probabilities of answers of a twig query. 
All the above work focused on the discussions of probabilistic XML data model and/or structured XML query, e.g., twig query.
The most closely related work is \cite{DBLP:conf/icde/LiLZW11} that proposed two algorithms to answer top-$k$ keyword queries over probabilistic XML data. However, compared with \cite{DBLP:conf/icde/LiLZW11}, in this work we propose a probabilistic inverted index that can be used to efficiently answer threshold keyword queries by reducing the computational cost of  unqualified nodes. In addition, we also take into account the relaxation (i.e., quasi-SLCA) of results for keyword search w.r.t. a threshold value while \cite{DBLP:conf/icde/LiLZW11} focused on the strict SLCA semantics of results.


There are some other work to discuss probabilistic index for query evaluation and/or data management. Although \cite{DBLP:conf/icde/SinghMPSH07} discussed probabilistic inverted index as ours, its data model is relational in which each tuple is associated with a probability value and all tuples are assumed independent. In our work, we built the probabilistic inverted index based on probabilistic XML data model with IND and MUX semantics. Another difference is that we answer keyword query while \cite{DBLP:conf/icde/SinghMPSH07} processes equality query.
Another work discussing probabilistic index is \cite{DBLP:conf/icde/VolkRHHL09} that first generates possible worlds and then cluster them based on probability values with a limited distance. The problem is that generating all possible worlds is very time-consuming in XML data. In our work, we avoided the generation of possible worlds.

\section{Conclusions}\label{sec:conclusions}
In this work, we first proposed and investigated the problem of finding \textit{quasi}-SLCA for PrTKQs over probabilistic XML data. And then we designed a PI index and analyzed the pruning features of PI index. Based on the lower and upper bounds computed from PI index, the proposed PI-based algorithm can quickly identify the qualified results and filter the unqualified ones. Our experimental results demonstrated the comparison of Baseline algorithm, PI-based Exact-computation Algorithm (PIEA) and PI-based Approximate-computation Algorithm (PIAA), which verified our motivation and approaches. 






\bibliographystyle{IEEEtran}

\bibliography{topkslca}  

\ifCLASSOPTIONcaptionsoff
  \newpage
\fi

\end{document}